\documentclass[prl,twocolumn,showpacs]{revtex4}
\usepackage{graphicx}
\usepackage{amssymb}
\usepackage{amsmath}
\usepackage{xspace}

\begin{document}

\title{Sparse polynomial space approach to dissipative quantum systems: \\ Application to the sub-ohmic spin-boson model}

\author{A.~Alvermann}
\author{H.~Fehske}
\affiliation{
Institut f\"ur Physik, Ernst-Moritz-Arndt-Universit{\"a}t
Greifswald, 17489 Greifswald, Germany }

\begin{abstract}
We propose a general numerical approach to open quantum systems
with a coupling to bath degrees of freedom. The technique
combines the methodology of polynomial expansions of spectral
functions with the sparse grid concept from interpolation theory.
Thereby we construct a Hilbert space of moderate dimension
to represent the bath degrees of freedom, which allows us to perform
highly accurate and efficient calculations of static, spectral and
dynamic quantities using standard exact diagonalization algorithms.
The strength of the approach is demonstrated for the phase transition,
critical behaviour, and dissipative spin dynamics in the spin boson model.
\end{abstract}

\maketitle

Whenever a small quantum object, such as an atom, molecule or quantum dot, is not perfectly isolated it couples to the degrees of freedom of its environment. 
In such an open quantum system the environment acts as a `bath' with which to exchange particles or energy with. A fermionic bath serves as a particle reservoir, while a bosonic bath accounts for dissipation~\cite{Wei99}.
Since the interest is only in the influence of the environment on the small quantum object, one may suspect that phenomenological descriptions of open quantum systems, e.g. by Lindblad equations for dissipative baths, are sufficient. 
But in general correlations between the quantum system and the bath evolve,
which can lead to strong renormalization as in the Kondo effect,
or determine the time evolution of observables in unexpected ways.
Simple phenomenological descriptions are obtained only within potentially unwarranted approximations such as weak coupling perturbation theory. 
To perform reliable computations for open quantum systems including correlations with the environment is a challenging problem for theoreticians.

A generic and important example  of an open quantum system is the spin-boson model~\cite{LCDFGZ67}. Its Hamiltonian
\begin{equation}\label{Ham}
H =  \frac{\Delta}{2} \sigma_x + \sum_i \lambda_i (b^+_i+b^{}_i) \sigma_z + \sum_i \omega_i b^+_i b^{}_i - \epsilon \sigma_z
\end{equation}
describes a spin-$1/2$ (with Pauli matrices $\sigma_i$) coupled to a bosonic bath of oscillators, whose dynamics is given by $H_B=\sum_i \omega_i b^+_i b^{}_i $.
The spin-boson coupling is specified by the spectral function
\begin{equation}
 J(\omega) = \sum_i \lambda_i^2 \delta(\omega-\omega_i) = \frac{\alpha}{2} \omega_c^{1-s} \omega^s \Theta(\omega_c-\omega) \;,
\end{equation}
with a power-law dependence $\propto \omega^s$ up to a cutoff frequency $\omega_c$ (we set $\omega_c=1$ in the examples below).
The spin-boson model shows rich physics beyond the dissipative spin dynamics at weak coupling.
In the sub-ohmic (ohmic) regime $s<1$ ($s=1$) the model
undergoes, for $\epsilon=0$, a quantum phase transition (QPT) from a non-degenerate groundstate with zero magnetization $m=\langle \sigma_z \rangle$ below a critical coupling $\alpha_c=\alpha_c(\Delta,s)$ to a two-fold degenerate groundstate with finite $m\ne 0$ for $\alpha>\alpha_c$.
The existence of the QPT is a consequence of the coupling of the spin to bosons at low frequencies, which may entirely suppress the spin dynamics.
In that respect the spin-boson model captures the renormalization aspect of Kondo physics.

Only few methods are capable of accessing the QPT in the sub-ohmic spin-boson model. 
Among them we find powerful numerical techniques such as the Numerical Renormalization Group (NRG)~\cite{BTV03,nrg} or Quantum Monte Carlo (QMC)~\cite{WRVB09}.
Prominently missing in the above enumeration
are techniques from the field of exact diagonalization (ED),
which are otherwise routinely used to yield highly accurate and unbiased results for strongly correlated systems~\cite{Da94}.
ED techniques require a finite-dimensional matrix representation of the model Hamiltonian.
Once the matrix is given the Lanczos algorithm allows for the calculation of the groundstate and a few excited states, while Chebyshev expansion techniques such as the Kernel Polynomial Method (KPM)~\cite{WWAF06} provide dynamic  properties, e.g. spectral functions at zero or finite temperature, as well as the time-evolution of the wavefunction~\cite{TK84}. 
The main obstacle against this procedure for the spin-boson model and open quantum systems in general is that a finite Hamiltonian matrix involves discretization of the continuous spectral density $J(\omega)$.
Naive discretization, i.e. the approximate replacement of $J(\omega)$ by a sum of $\delta$-peaks,
requires either a very large number of bosonic orbitals, which
leads to matrices beyond any accessible size, or obtains results spoiled by discretization artefacts.

The sparse polynomial space representation (SPSR) we propose in this Letter overcomes the ED restriction.
It avoids the discretization of the bath spectral function $J(\omega)$ and constructs a Hilbert space of moderate dimension to represent continuous bath degrees of freedom with high resolution.  
In that way the SPSR extends the Chebyshev space method developed in Ref.~\cite{AF08},
and it becomes possible to perform efficient and accurate calculations for open quantum systems using ED algorithms.
As a non-trivial example we analyse the QPT and the dissipative spin dynamics in the spin-boson model.

The Hilbert space of the Hamiltonian~\eqref{Ham} is the tensor product of the spin space 
$\mathbb{C}^2$ with the bosonic Fock space $\mathcal{B}$.
To set up the SPSR for $\mathcal{B}$ we proceed in three steps:
we (i) parametrize multiple bosonic excitations through symmetric wavefunctions as in first quantization, 
(ii) expand these wavefunctions into orthogonal polynomials,
(iii) select a sparse subspace of the polynomial space.

\begin{figure}
\includegraphics[width=\linewidth]{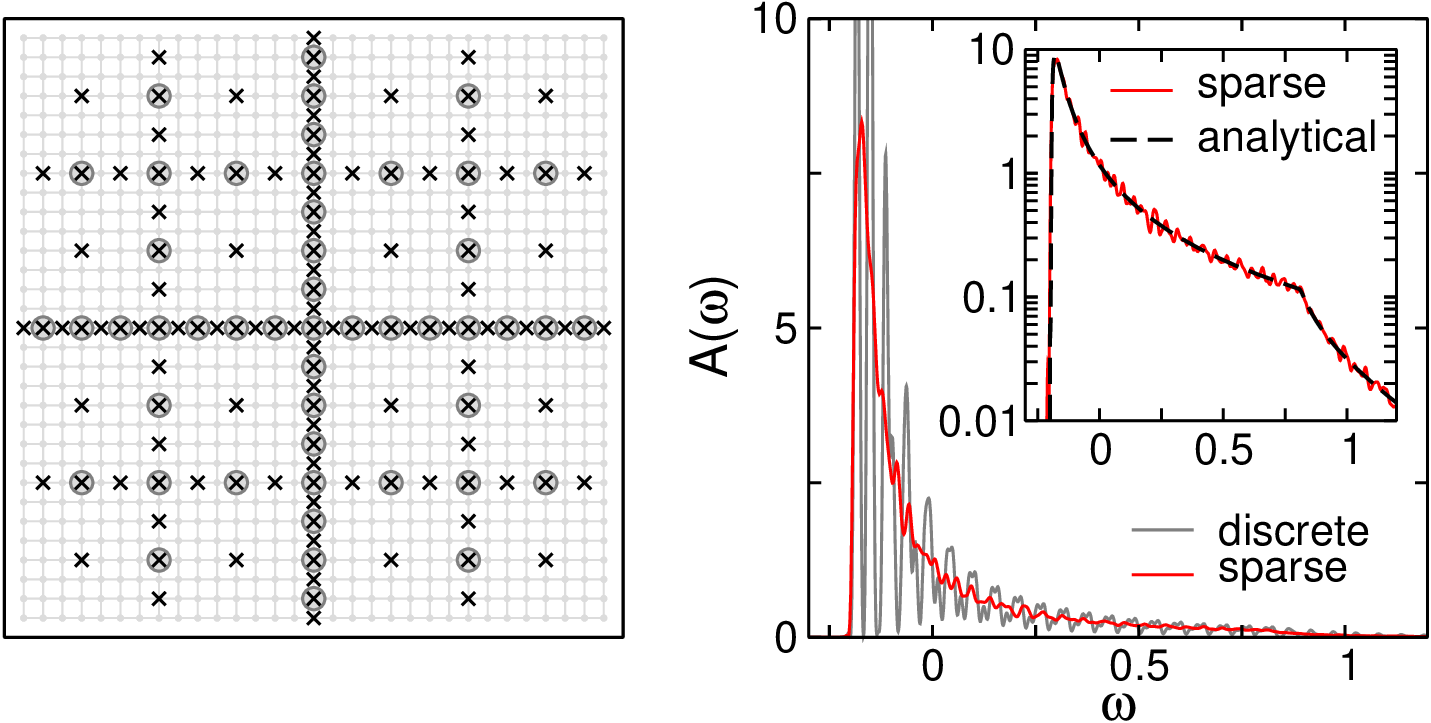}
\caption{(Color online) Left panel: Two-dimensional sparse grid of level $N_g=3$ (circles) and $N_g=4$ (crosses).
Right panel: Spectral function $A(\omega) = \langle\uparrow;\mathrm{vac}| \delta[\omega-H]|\uparrow;\mathrm{vac}\rangle$
for $\Delta=0$, $s=0.5$, $\alpha=0.2$ calculated using KPM. Keeping up to $N_b=6$ bosons, the SPSR to level $N_g=10$ contains $129284$ states.
The comparable discrete grid ($134596$ states)  
contains only $N_p=18$ orbitals  (here at equidistant energies $\omega_i$).
}
\label{Fig1}
\end{figure}

For step (i) we fix an (unnormalized) density of states $D(\omega)=\sum_i \delta(\omega-\omega_i)$ on $[0,\omega_c]$, which must be a smooth function for a continuous spectral function $J(\omega)$.
In our numerics we use $D(\omega) \propto (1-x^2)^{-1/2}$ 
with $x=(2\omega/\omega_c-1) \in (-1,1)$, which will lead to Chebyshev polynomials in step (ii).
In first quantization any $n$-boson state $|\psi_n\rangle$ is represented by a totally symmetric wavefunction $\psi_n:   [0,\omega_c]^n \to \mathbb{C}, \vec{\omega} \mapsto \psi_n(\vec{\omega})$.
Here, the argument $\vec{\omega}=(\omega_1,\dots,\omega_n)$ of the wavefunction gives the boson energies.
We find that $H_B$ multiplies the value $\psi_n(\vec{\omega})$ to argument $\vec{\omega}$ by the total energy $\sum_i \omega_i$.

To express the Hamiltonian Eq.~\eqref{Ham} in our calculations we further need the operators 
$b^{(+)}=\sum_i \lambda_i b_i^{(+)}$.
These are bosonic operators up to normalization, since $[b,b^+] = \sum_i \lambda_i^2 = \int d\omega J(\omega)$.
We choose the function $\lambda(\omega)$ such that $J(\omega) = \lambda(\omega)^2 D(\omega)$, or $\lambda(\omega_i) = \lambda_i$ in comparison to Eq.~\eqref{Ham}.
Then the single-boson state $b^+|\mathrm{vac}\rangle$ is represented by the wavefunction 
$\psi_1(\omega)=\lambda(\omega)$.
Straightforward calculations show how to obtain the wavefunctions of any state  $b^{(+)}|\psi_n\rangle$.  
We note exemplarily, that for a single boson state $|\psi_1\rangle$ with wavefunction $\psi_1(\omega_1)$, the state $b^+|\psi_1\rangle$ has wavefunction $\psi_2(\omega_1,\omega_2) = (\psi_1(\omega_1) \lambda(\omega_2)+\lambda(\omega_1)\psi_1(\omega_2))/\sqrt{2}$, while $b|\psi_1\rangle$ is the scalar $\int d\omega D(\omega) \lambda(\omega) \psi_1(\omega)$.

For step (ii), 
note that the scalar product of wavefunctions is given by
\begin{equation}\label{ScalProd}
(\psi_n,\phi_n) = \int_{[0,\omega_c]^n}  \prod_i D(\omega_i) d\omega_i \; \psi_n^*(\vec{\omega})  \phi_n(\vec{\omega}) \;.
\end{equation}
Therefore we choose polynomials $P_m$ of degree $m$ for $m\ge 0$ subject to the orthonormality condition
 \begin{equation}
 \int_0^{\omega_c} d\omega D(\omega) P_l(\omega) P_m(\omega) = \delta_{lm} \;.
\end{equation}
For the above choice of $D(\omega)$, the $P_m$ are scaled and shifted Chebyshev polynomials. 
Any wavefunction $\psi_n(\vec{\omega})$ has an expansion
\begin{equation}\label{waveexp}
 \psi_n(\vec{\omega}) = \sum_{\vec{m}} \psi_{\vec{m}} \prod_{i=1}^n P_{m_i}(\omega_i) 
\end{equation}
in that complete polynomial function system.
Therefore the multi-indices $\vec{m}$ enumerate the elements of an orthonormal basis of $\mathcal{B}$.
Instead with the wavefunction $\psi_n(\vec{\omega})$ we can calculate with 
the (totally symmetric) coefficients $\psi_{\vec{m}} = \int_{[0,\omega_c]^n}  d\vec{\omega} \prod_i D(\omega_i) P_{m_i}(\omega_i) \psi(\vec{\omega})$.

Generally, orthogonal polynomials $P_m$ obey a three-term recurrence~\cite{Gau04} of the form
$P_{m+1}= (a_m \omega -b_m) P_m - c_m P_{m-1}$.
Owing to this recurrence the multiplication with $\sum_i \omega_i$ occurring for the operator $H_B$ affects the coefficients $\psi_{\vec{m}}$ only with index shifts by at most $\pm 1$.
To obtain the operator $b^{(+)}$ we use the expansion
$\lambda(\omega) = \sum_m \lambda_m P_m(\omega)$
and find, e.g., that $b^+|\mathrm{vac}\rangle$ has coefficients $\psi_m = \lambda_m$.
Similarly,
for a single boson state $|\psi\rangle$ with coefficients $\psi_m$, the state $b^+|\psi\rangle$ has coefficients $\psi_{(m_1,m_2)} = (\psi_{m_1} \lambda_{m_2}+\lambda_{m_1}\psi_{m_2})/\sqrt{2}$, while $b|\psi\rangle$ is the scalar $\sum \lambda_m \psi_m$.

The bosonic Fock space and all relevant operators are now expressed by simple operations on a polynomial space.
To prepare step (iii) notice that the selection of a finite dimensional subspace containing all polynomials up to degree $N_p$ is equivalent to naive discretization of $J(\omega)$,
with $N_p+1$ energy levels $\omega_i$ given as the zeroes of  $P_{N_p+1}(\omega)$.
This discrete grid requires $(n+N_p)!/(n! N_p!)$ coefficients to represent an $n$-boson state.
To overcome the `curse of dimension' expressed by the exponential growth of the binomial with $n$ 
we resort to the concept of sparse grids~\cite{Grid} from interpolation theory.

An $n$-dimensional sparse grid of level $N_g$ is a subset of the Cartesian grid with $(2^{N_g}-1)^n$ points (see Fig.~\ref{Fig1}).
With the sparse grid comes an interpolation formula
that assigns a polynomial to given function values at the sparse grid points.
This interpolation has the property that functions of bounded variation are approximated with high accuracy although the number of points is significantly smaller than in the Cartesian grid.
For our purposes we do not access the points of the sparse grid directly.
Instead we note that the sparse grid interpolation formula is exact for a polynomial subspace of the full function space.
Exactly this sparse polynomial space is selected for the SPSR.
Assigning to a polynomial of degree $m$  a logarithmic `cost' $co[m] = \lfloor \log_2  (m+1) \rfloor$ 
(rounding down to an integer), we keep in step (iii) all polynomial basis states with multi-indices that satisfy
\begin{equation}
 co_n[\vec{m}] = \sum_{i=1}^n co[m_i]  \le  N_g \;.
\end{equation}
For a single bosonic excitation ($n=1$) the SPSR of level $N_g$ contains all polynomials with degree $m < 2^{N_g}-1$.
For $n>1$, the SPSR contains only a small fraction of all polynomials,
discarding those combinations where many polynomials have large degree.
The motivation is that for multiple excitations 
the fine structure of the energy distributions among the various excitations
becomes less important than for few excitations.
Although the motivation is related to Monte Carlo sampling of the state space,
the SPSR is deterministic without statistical error.
Note further that, increasing $N_g$, the SPSR is truly variational for the groundstate.

In Fig.~\ref{Fig1} the SPSR is compared to a discrete grid for the calculation of a spectral function. 
The discrete grid calculation is dominated by artefacts introduced by the inescapable restriction to a small number of orbitals.
It is evident that the SPSR succeeds:
Multiple bosonic excitations for continuous bath degrees of freedom
are accurately represented with a moderate effort.
Note that the SPRS resolves the jump discontinuity of $A(\omega)$ at the groundstate energy $E_{sh} = \int d\omega J(\omega)/\omega$, and has uniform resolution over the full energy range.

\begin{figure}
\includegraphics[width=\linewidth]{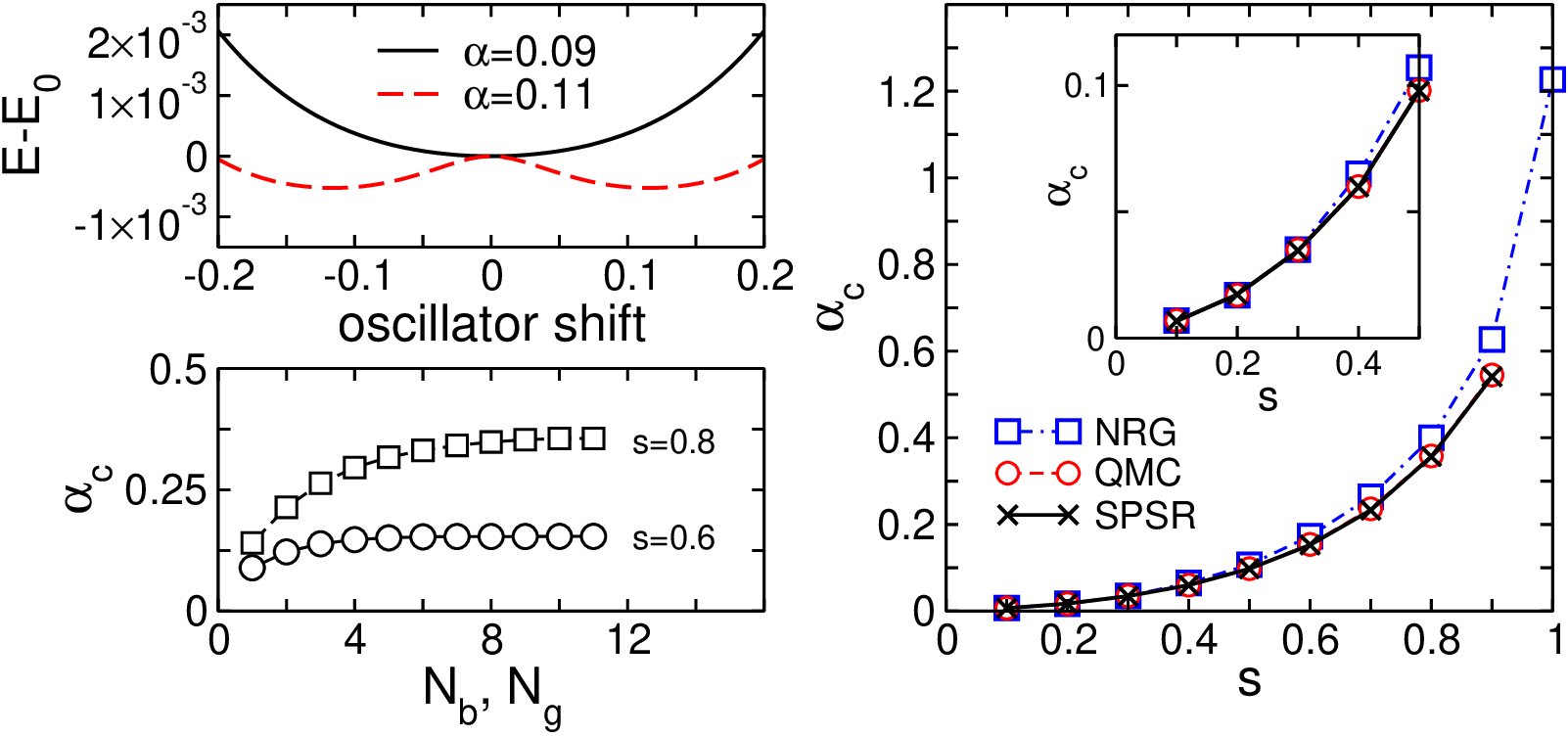}
\caption{(Color online) Left panels: Groundstate energy $E$ as a function of oscillator shift $\delta$; convergence of critical coupling $\alpha_c$ with increasing Hilbert space size $N_g$, number of bosons $N_b$. 
Right panel: Phase diagram of the sub-ohmic spin-boson model for $\Delta=0.1$, i.e. $\alpha_c$ as a function of $s$, in comparison to QMC/NRG data taken from Ref.~\cite{WRVB09}.
}
\label{Fig2}
\end{figure}

\begin{figure}
\includegraphics[width=\linewidth]{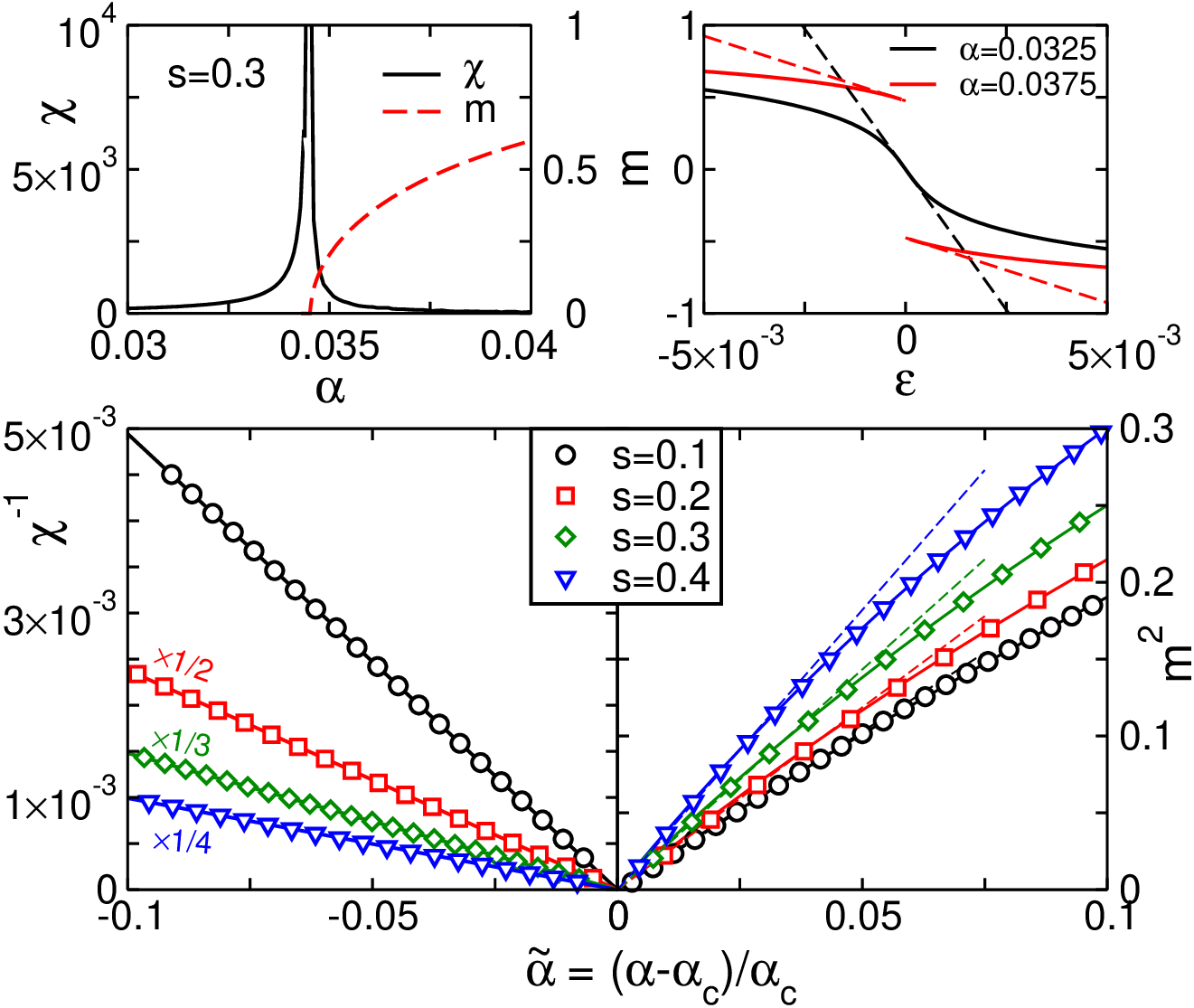}
\caption{(Color online) Upper left panel: Susceptibility $\chi$ and magnetization $m$ as function of $\alpha$, for $s=0.3$ and $\Delta=0.1$.
Upper right panel: Magnetization $m$ as function of external field $\epsilon$, still for $s=0.3$ and $\Delta=0.1$. The dashed straight lines indicate the slope of $m$ for $\epsilon\to 0$, which determines $\chi$.
Lower panel: Critical behaviour of $\chi$ and $m$ close to the phase transition, for $\Delta=0.1$ and $s<1/2$. The curves for $\chi$ are multiplied with the indicated factors for better visibility. The straight lines indicate the critical behavior for $\tilde{\alpha}=(\alpha-\alpha_c)/\alpha_c \to 0$. The solid curves on the right show a fit to the ansatz $m\propto \tilde{\alpha}^\beta (1+O(\tilde{\alpha}))$, which results in the critical exponent $\beta=1/2$ within numerical accuracy.}
\label{Fig3}
\end{figure}

\begin{figure}
  \includegraphics[width=\linewidth]{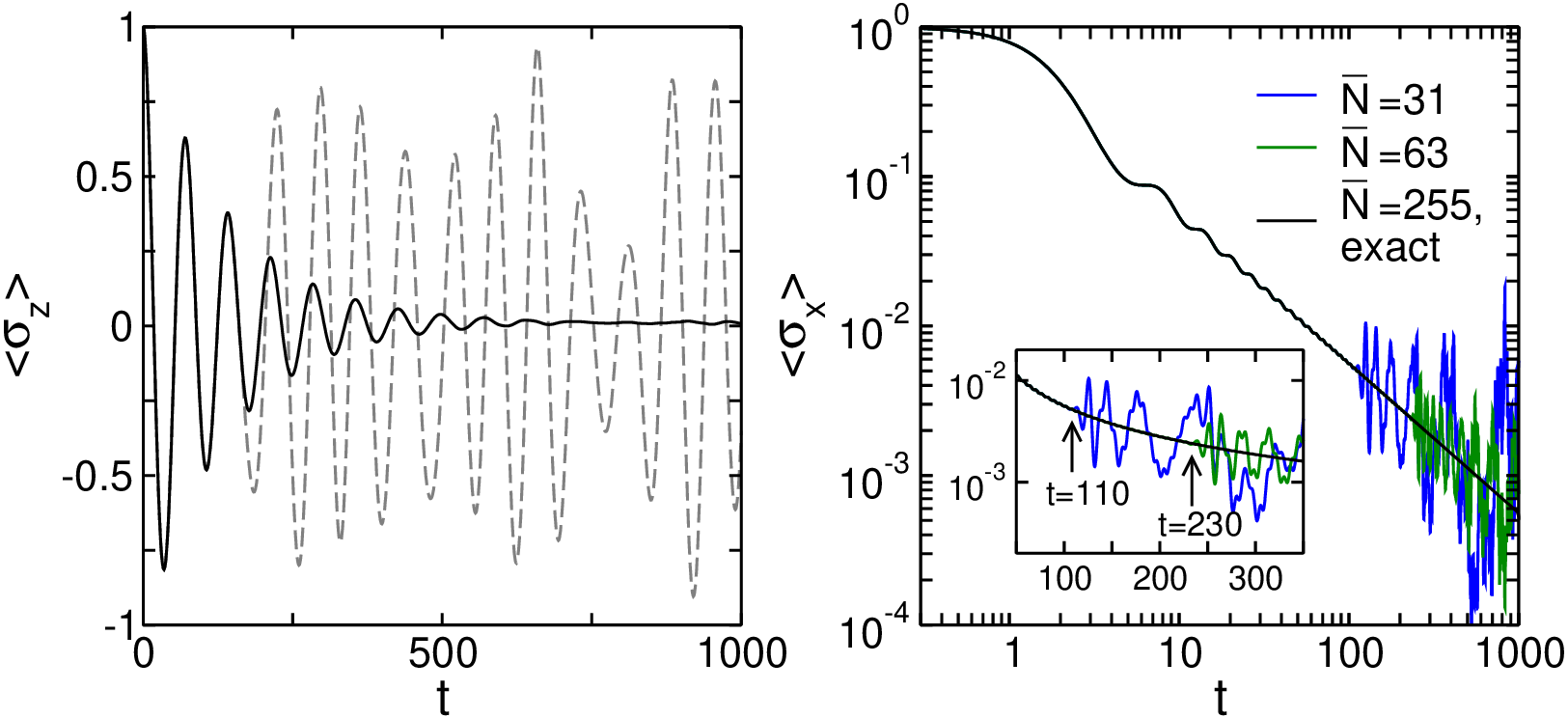}
\caption{(Color online) Left panel: Magnetization $m(t)=\langle \psi(t)| \sigma_z |\psi(t) \rangle$
for $\Delta=0.1$, $s=1$, $\alpha=0.05$ (with $N_g=12$, $N_b=6$). For $t<0$ the system is prepared as a spin-up state with relaxed bosonic bath. The dashed curve shows the result from a discrete grid ($N_p=25$).
Right panel: Decay of $\langle\sigma_x(t)\rangle$ for $\Delta=0$, $s=1$, $\alpha=0.5$.
For $t<0$ the system is prepared as a spin singlet in the bosonic vacuum. 
We calculate $\sigma_x(t)$ in the Heisenberg picture using a polynomial representation of operators. Already with $\bar{N}=255$ polynomials (corresponding to $N_g=8$) the numerical and analytical results match up to $t=1000$.
}
\label{Fig4}
\end{figure}

To put the SPSR to a severe test we calculate the phase transition in the sub-ohmic spin boson model.
A NRG study~\cite{BTV03} of the QPT 
obtained for $s<1/2$ critical behaviour incompatible with a mean-field transition expected 
from the quantum-classical mapping to the Ising spin chain with long-range interactions~\cite{LCDFGZ67}.
Using QMC the authors recently corrected these findings~\cite{WRVB09},
confirming  a mean-field transition.
Apparently, the NRG calculations of the critical behaviour suffered from a subtle error inherent to the renormalization scheme.
In light of this controversy we use the SPSR to analyse the QPT independent of previous calculations.

The QPT is best detected using the relation $\langle b_i^+ + b_i\rangle = - 2 (\lambda_i/\omega_i) m$ between oscillator shift and magnetization in the groundstate.
We therefore consider the Hamiltonian
\begin{equation}
 \tilde{H}(\delta) = H + \delta \sum_i \lambda_i (b_i^+ + b_i) + 2 \delta E_{sh} \sigma_z + \delta^2 E_{sh}  \;,
\end{equation}
where the oscillator shift is introduced via the unitary transformation $U(\delta)= \exp[\delta \sum_i (\lambda_i/\omega_i) (b_i^+ - b_i) ]$.
In a certain sense $U(\delta)$ prepares a classical mean-field state, 
while the quantum fluctuations are captured by the SPSR.
Of course, the true groundstate energy $E(\delta)$ of $H(\delta)$ is independent of $\delta$.
But the SPSR becomes optimal if the oscillator shift, hence the average boson number, is small.
Consequently, the numerical $E(\delta)$ is minimal at finite (zero) $\delta$
if the true groundstate has finite (zero) magnetization (see Fig.~\ref{Fig2}).
From $E(\delta)$, calculated e.g. with the Lanczos algorithm, 
we obtain the critical coupling $\alpha_c$ by simple bisection.
Increasing the number of states in the SPSR the numerical values converge to the true $\alpha_c$ (lower left panel), which in turn yields the phase diagram (right panel).
In Fig.~\ref{Fig3} we show the groundstate magnetization $m$ and the susceptibility $\chi = \lim_{\epsilon\to 0}(\partial m/\partial \epsilon)$.
The critical behaviour of the two quantities clearly confirms a mean-field transition for $s<0.5$ 
with $m \sim (\alpha-\alpha_c)^{1/2}$ and $\chi \sim (\alpha_c-\alpha)^{-1}$ (Fig.~\ref{Fig3}, lower panel). 
Note that probing for finite $m$ or the divergence of $\chi$ is an alternative to the above QPT criterion.
The obtained values for $\alpha_c$ agree with each other (cf. Fig.~\ref{Fig2} and Fig.~\ref{Fig3} for $s=0.3$), but
the above criterion is easier evaluated within the numerics,
while e.g. $\chi$ is obtained as a derivative.

The analysis of the QPT demonstrates that the SPSR carries the unique virtues of ED techniques
over to open quantum systems.
Physical properties are found by the direct calculation of the corresponding observables.
No scaling or extrapolation involving additional assumptions are required,
no method specific quantities enter the discussion.
The computational effort is moderate, ranging from a few minutes to hours on  standard PCs for the given results.
Concerning their quality, our phase diagram is in perfect agreement with QMC and, taking the logarithmic NRG discretization into account, also with NRG.
Our data for the critical behaviour confirm the QMC data, 
extrapolated to zero temperature.
Here we can read off the critical behaviour directly from the numerical values.

ED techniques have the overall advantage that, once the Hamiltonian matrix is given,
almost any associated quantity can be obtained with high precision.
Since the major interest is in the dynamics of open quantum system,
we finally give a single example for the dissipative spin dynamics at weak coupling.
Efficient time evolution with Chebyshev techniques~\cite{TK84,AF08} gives the magnetization
as a function of time (Fig.~\ref{Fig4}, left panel).
The curves are in perfect agreement with the results from time-dependent NRG~\cite{AS06} and unitary perturbation theory~\cite{HK08}.
A special feature of our calculation is that it requires no additional damping, and no averaging over different bath discretizations.	
This results from the superior resolution provided by the SPSR even for multiple bosonic excitations. 
Although we have demonstrated that the SPSR is not restricted to weak coupling
the time evolution close to the QPT deserves a careful examination that we postpone to a future publication. 
To indicate the potential as the final example we show the decay of $\langle\sigma_x(t)\rangle$ for $\Delta=0$.
For a finite number of polynomials the numerics exactly reproduces the analytical result, but only up to a finite time. With more polynomials that time can be easily made very large (Fig.~\ref{Fig4}, right panel).

In conclusion, 
we introduced the SPSR as a novel approach to static and dynamic properties of open quantum systems.
The SPSR involves a highly accurate representation of continuous bath degrees of freedom,
which is based on the sparse grid concept applied to polynomial expansions of wavefunctions.
It avoids the discretization artefacts that previously prevented the application of powerful ED techniques 
in presence of a bath.
We demonstrated the strength of the SPSR for the QPT in the sub-ohmic spin-boson model, where we confirm the quantum-to-classical mapping for $s<1/2$, and for the dissipative spin dynamics.
Despite its current early state of development we believe to have presented the SPSR as a serious alternative to more established methods.
An important issue for future work is the extension to fermionic baths,
which is possible using antisymmetrized wavefunctions in step (i) of our construction.
The effectiveness of SPRS in that case has yet to be assessed.

We acknowledge financial support by DFG through SFB 652.


\end{document}